\documentstyle[preprint,revtex]{aps}
\begin{document}

\draft
\preprint{LA-UR-93-2756}

\begin{title}
 On superconducting instability in non-Fermi liquid: scaling approach
\end{title}

\author{A. V. Balatsky}
\begin{instit}
 Theoretical Division\\ Los
Alamos National Laboratory, Los Alamos, NM 87545 and\\ Landau
Institute for Theoretical Physics, Moscow, Russia
\end{instit}

\receipt{August 5, 1993}
\begin{abstract}
The superconducting instability in a non-Fermi liquid in $ d>1$ is considered.
For a particular form of the  single particle spectral function with
homogeneous scaling $A(\Lambda k, \Lambda \omega) = \Lambda^{\alpha} A(k,
\omega)$ it is shown that the pair susceptibility is also a scaling function of
temperature with power defined by $\alpha$. We find three different regimes
depending on the scaling  constant. The BCS result is recovered for $\alpha =
-1$ and it corresponds to a marginal scaling of the coupling constant. For
$\alpha > -1$  the superconducting transition happens above some critical
coupling.  In the opposite case of $\alpha < -1$ for any fixed coupling  the
system undergoes a transition at low temperatures. Possible implications for
theories of high-$T_c$ with a superconducting transition driven by the
interlayer Josephson tunneling are discussed.

\end{abstract}


The question of non-Fermi liquid  behavior in higher dimensions ($d>1$) has
been addressed recently in the context of possible description of the normal
state of high temperature superconductors \cite{MFL,And1,S}. The solution of
this question , apart from  understanding  the normal state of high-T$_c$
materials also might lead to a better understanding of the  possible
superconducting and density wave instabilities of the non-Fermi liquid state.

 Recently a  general class of systems with a notrivial scaling coefficient has
been proposed in
\cite{S} as a possible description of the normal state of the  cuprate layers.
It was pointed out that for the  non-Fermi liquid form  of $A(k, \omega)$
interlayer single particle tunneling is  strongly suppressed and only Josephson
tunneling is relevant at low enegries. This Josephson interlayer tunneling
enhances the effective superconducting transition temperature, driven by {\em
intralayer} attraction. Although the non-Fermi liquid behavior was assumed
inside each layer in \cite{S}, the superconducting transition was considered
within the BCS theory and the in-plane  pair susceptibility was taken in the
BSC form $\chi_{pair} = th(\epsilon_k/2T) /\epsilon_k $. This choice was argued
to give qualitatively correct answer for the large  enough transition
temperatures. However in general  the true pairing susceptibility in a
non-Fermi liquid should be used to identify the superconducting instability of
the normal state.


In this note we will consider the superconducting instability of a non-Fermi
liquid in $d>1$, using the same set of assumptions made in \cite{S}. Let us
leave the most important and interesting question of the origin of  the
breakdown of Fermi liquid unanswered and consider the effect of attraction
between  excitations in a non-Fermi liquid  and conditions  under which this
attraction will lead to   superconductivity.  We will argue that to be
consistent one has to calculate the pair susceptibility using non-Fermi liquid
Green's functions. It is shown  that for the non-Fermi liquid behavior,
characterized by the vanishing quasiparticle residue $Z_{\omega}$ at low
frequency, the pair susceptibility is lower then  in the Fermi liquid and is of
a non-BCS form for general values of $\alpha$. This results in a  {\em
qualitative} difference with the  BCS instability: we find  a  critical
coupling for the superconducting transition for the case of a Luttinger liquid
behavior, bel!
ow which the system remains normal

 down to $T = 0$  \cite{later}.

 As is well known,  in the  Luttinger liquid  in $d = 1$ we always have
competing density waves and superconducting interactions coming from the same
terms in the Hamiltonian. These interactions do not produce true long range
order but they lead to  power law correlators. Presumably at $d>1$ interaction
will lead to a true instability in one of the channels. The artificial nesting
will disappear in $d>1$ (except in some special cases) and we can consider
different channels independently. We  will use the ladder approximation  as was
used in \cite{S}, ignoring parquet effects.

The absence of a  microscopic description of a non-Fermi liquid for $d > 1$
leads us to consider the phenomenology of this state, analogous to the ideas
proposed in \cite{MFL,S}. We assume that this state  supports  single particle
excitations which obey Fermi statistics with a  Green's function:
\begin{equation}
G(k, \omega_n) = \int_{-\infty}^{+\infty} {A(k, x)\over{x - i\omega_n}} dx
\end{equation}

With the spectral function obeying the  following homogeneous equation:
\begin{eqnarray}
A(\Lambda k, \Lambda \omega) = \Lambda^{\alpha} A(k, \omega)
\end{eqnarray}

with some scaling coefficient $\alpha$.
 We assume an isotropic dispersion hereafter and only the  magnitude of the
momentum, counted from the Fermi momentum $k_F$, enters into the spectral
function. This scaling form  taken over the whole frequency range violates the
spectral sum rule, except for $\alpha = -1$. Indeed $  \int_{-\infty}^{+\infty}
A(k, \omega) d \omega = \Lambda^{\alpha + 1} \int_{-\infty}^{+\infty} A(k,
\omega) d \omega  = 1$, where the second equation is obtained by rescaling
$\omega \rightarrow \Lambda \omega$. We consider this scaling form of the
spectral function  to be valid only for the low energy part of the spectrum.

 The simplest example of the scaling law of this type is a Fermi liquid
spectral function:
\begin{eqnarray}
A(k, \omega) =  Z \pi \delta(\omega - v_Fk), \ \ \ \alpha  = -1
\end{eqnarray}
and the scaling coefficient follows immediately from the fact that
$\delta(\Lambda a) = \Lambda^{-1} \delta(a)$. Another example of a system with
$\alpha = -1$ , up to irrelevant logarithmic factors, is a marginal Fermi
liquid with $A(k, \omega)  = -2 Im( \omega - v_Fk - \gamma[\omega log \omega +
i \pi \omega])^{-1}$ , proposed in \cite{MFL}. The case of general scaling is
realized for a Luttinger liquid Green's function, e. g. $G(k, \omega) \sim
(\omega^2 - v^2_c k^2)^g /(\omega - v_c k)$ with $\alpha = 2g - 1 > -1$.

 Here we will show that under the  scaling assumptions of Eq.(2) the in-plane
susceptibility will be a non-BCS form  and subsequently the theory of the
superconducting transition, driven by an in plane attraction, will be
different from BCS. The main difference will come from the fact that for
general scaling exponent $\alpha$ the temperature dependence of the pair
susceptibility will be a power law with the  power dependent on $\alpha$. For
the physically  interesting case of $\alpha > -1$, which leads to  a Luttinger
liquid behavior in $d>1$,  we find a critical coupling,  above which  the
superconducting transition  is possible.

 We now consider a small s-wave attractive interaction:
\begin{eqnarray}
H_{int} = - V \int d{\bf r} c^{\dag}_{\uparrow}({\bf r})
c^{\dag}_{\downarrow}({\bf r}) c_{\downarrow}({\bf r}) c_{\uparrow}({\bf r})
\end{eqnarray}
In the ladder approximation within the weak coupling theory we have for a
critical $T_c$:
\begin{eqnarray}
1 = g T\sum_{n, {\bf k}} G(k, \omega_n) G(-k, -\omega_n)
\label{g}
\end{eqnarray}
where $g = -V N_0$. Using the spectral representation Eq.(1) and rescaling
variables as $\overline{x} = x \beta, \overline{y} = y \beta, v_F\overline{k} =
v_Fk \beta, \beta = 1/T$ in the integral in this equation we find:
\begin{equation}
1 = g \beta^{-2(1 + \alpha)} \int_{0}^{\omega_o \beta} d \overline{\xi}
F(\overline{k})
\label{int}
\end{equation}
where,
\begin{equation}
F(\overline{k}) = \int \int d\overline{x} d\overline{y} A(\overline{k},
\overline{x}) A(\overline{k}, \overline{y}) (th \overline{x}/2 + th
\overline{y}/2)(\overline{x} + \overline{y})^{-1}
\label{f}
\end{equation}
with $\omega_0$ being the upper cut off. In deriving  Eq.(\ref{int}) we assume
that the scaling form of $A(k, \omega)$ is valid at {\em all} temperatures
above $T_c$ and that the density of states $N_0$ is invariant under scaling,
which is certainly true in $2d$ and it holds in general upon ignoring
particle-hole asymmetry. The selfconsistent treatment of the gap equation below
$T_c$ is a more complicated matter due to a possible breakdown of the scaling
in Eq.(2) and we will not consider it here.

An important comment is in order here.  As was mentioned above,  the scaling
form of the spectral function can not be valid in the whole frequency range.
We consider the case when $A(k, \omega) = \omega^{\alpha} f(\omega/k) +
A_{inc}(k, \omega)$ with normalization $\int_{-\infty}^{+\infty} dx f(x) = 1$.
The first term is  a ``scale invariant" part of the spectral function, the
second term violates scaling and is small at low frequency. One can show that
at $T \sim \Lambda \rightarrow 0$ the dominant part in the spectral sum comes
from  $A_{inc}(\Lambda k, \Lambda \omega)$. Thus in the integral in
Eq.(\ref{int}) terms, containing $ A_{inc}( k, \omega)$, will dominate at  $T
\sim \Lambda \rightarrow 0$. However, for the general case of  not quite small
$\Lambda  \ ,  \ 0 < \Lambda < 1$, we assume that the scale invariant part
provides a major contribution, what allows us to get Eq.(\ref{f}).  This also
means that Eqs.(\ref{int}, \ref{f})
are invalid at sufficiently low temperatures.

 We generally can consider two possibilities: 1) the integral of
$F(\overline{k})$ is  convergent at the upper limit and 2) the integral is
divergent at the upper limit. In the first case we can safely put the  upper
limit to infinity. And the answer for the  dependence of $T_c$ on $g$ is {\em
universal}. The simplest example of this sort will be, say, power law decaying
$F(\overline{k})$. In the second case the dependence of the integral on the
upper cut off is crucial and obviously the result depends on the
{\em specific} form of $F(\overline{k})$ \footnotemark  \footnotetext{ I  am
indebted to J.R. Schrieffer for pointing out this possibility}. We can still
estimate the integral if we would assume some asymptotic form for the  integral
on the upper limit. However this would require extra input apart from the
scaling coefficient $\alpha$. The simplest example of this sort is the BCS case
with $ F(\overline{k}) =  th(\overline{x}/2)/\overline{x}$ which leads to a
logarithmically divergent integral in Eq.(\ref{int}).

Let us  concentrate now on the first case, when the answer is universal and
independent of the detailed form of $ F(\overline{k})$. Then the momentum
integral is trivial and equals   some number which can be incorporated into the
definition of the coupling constant. The final scaling law for coupling
constant $g$ {\em vs} $T_c$ follows immediately from Eq.(\ref{int}):
\begin{equation}
1  =  g (W/T)^{-2(1 + \alpha)}
\label{main}
\end{equation}
where $W$ is the energy scale, within which the scaling form of the spectral
function is assumed. This equation is the main result of this note. It gives
the dependence of the critical temperature $T_c$ on the coupling constant $g$.
It is useful to write it in the form of a scaling equation:
\begin{eqnarray}
\partial g/ \partial log \beta = 2(1 + \alpha) g + O(g^2)
\label{scaling}
\end{eqnarray}

We want to discuss now  possible regimes for  $g$ {\em vs} $T_c$ for different
coefficients $\alpha$. From Eqs.(\ref{main}, \ref{scaling}) follows that the
critical value $\alpha_c = -1$ precisely corresponds to a Fermi liquid case,
when the BCS equation yields a $log (\omega_0/T)$ in Eq.(\ref{int}). The Fermi
liquid superconducting instability is a marginal case with a quadratic
$\beta$-function.

For  $\alpha > \alpha_c$ the solution of the $T_c$ equation, Eq.(\ref{main}),
is $T_c/W= ({1\over{g}})^{{1\over{2(1 + \alpha)}}}$. From this it follows that
the  smaller  $T_c$ is the larger is coupling constant required to produce the
instability. For  any fixed coupling constant at $T/W <
({1\over{g}})^{{1\over{2(1 + \alpha)}}}$ the state will remain normal and the
pair susceptibility is finite. The same result also follows from the
$\beta$-function in Eq.(\ref{scaling}) which indicates that for lower
temperatures one has to go to higher coupling to produce an instability. We
find that there always is a critical coupling
\begin{equation}
g_{min} \sim (W/W')^{2(1 + \alpha)} \sim 0(1)
\end{equation}
  for the  superconducting transition, $W'$ is some energy scale of the order
of $W$. One can show that for the case 2) it is true as well.

 This result leads immediately to a question of {\em reentrant} superconducting
behavior at high temperatures, see Fig.1. The slope of the function $g_c(T)$ is
negative at low temperatures. It would correspond to an {\em irrelevant}
coupling at low temperatures. However one  can not exclude the possibility of a
superconducting transition  at high temperatures which   then reenters the
normal phase at low temperatures $T/W < ({1\over{g}})^{{1\over{2(1 +
\alpha)}}}$.

It should be  stressed  that these results depend on the   strong assumption of
the convergence  of the  integral in Eq.(\ref{int}). Otherwise the dependence
on the upper cut off can change the function $g_c(T)$. We found   examples of
this sort, where  this dependence even for $\alpha > \alpha_c$ leads to a
nonreentrant phase diagram with $g_c(T)$ monotonically increasing.  In these
cases  the transition is still possible only at a coupling constant greater
then some critical value, as is natural for a noninfrared theory.

For the opposite case $\alpha < \alpha_c$ the transition temperature is always
nonzero for any coupling $g$ with $g_c(T)$ monotonically increasing.  This
regime corresponds to a {\em relevant}  superconducting coupling. The case
$\alpha < \alpha_c$ appears to have no analog in a $1d$ Luttinger liquid and
we will not discuss it here.

In the recent theory \cite{S} the non-Fermi liquid behavior in the spectral
form Eq. (2) with $\alpha > 0$ was considered as a model of a normal state
inside each layer in high-$T_c$ materials.  Our results shows that the
superconducting transition in each layer will require some critical coupling,
for the physically most interesting case of $\alpha > -1$, in order to have a
superconducting transition {\em inside}  each layer.  This result is {\em
qualitatively} different from the BCS pairing theory in which for any coupling
$T_c > 0$.

There  is one interesting   possibility to consider when the  superconducting
coupling inside each layer is below critical and Josephson tunneling between
different $Cu-O$  planes, enhancing  attraction in each layer, provides this
critical coupling. The simplest example is a model of two planes with non-Fermi
liquid behavior, coupled via a single particle tunneling matrix element
$t_{\perp}$. If one has an in-plane coupling  $g < g_{min}$ then strong enough
 $t_{\perp}$ will produce superconductivity in these two planes simultaneously.
Whereas, if to take $t_{\perp} = 0$, each plane separately will become
superconducting at much lower temperature. At this point we go beyond the
considered  model and assume that at low enough temperatures Fermi liquid
behavior sets in and a small in-plane attraction will produce a BCS
instability.

\underline{Conclusion} We considered  the superconducting transition in the
non-Fermi liquid with special scaling properties Eqs. (1,2). In general the
pair susceptibility is found to be of non-BCS form. For the scaling coefficient
$\alpha  = -1$ we recover the  BCS results with the marginal scaling of the
coupling constant. For the most interesting case of the Luttinger liquid
Green's functions  with $\alpha > -1$ we find that superconducting transition
requires a threshold value of the coupling constant and is {\em qualitatively}
different from the BCS case where the instability is caused by arbitrarily
small coupling.

 In general the results depend on the upper cut off in the momentum sum.
However, for a special case of the convergent momentum sum the  results are
universal and depend  on the index $\alpha$. In the opposite case of $\alpha <
-1$ the normal ground state is {\em always} unstable for all values of  the
coupling constant.

\underline{Acknowledgments} I am grateful to E. Abrahams, K. Bedell, D. Coffey,
 P.A.  Lee, P. Littlewood and J.R. Schrieffer and Z.Q. Wang for useful
discussions. I am particularly grateful to  S. Chakravarty for  discussions on
the role of  scaling and interlayer tunneling in \cite{S}.  This work was
supported by J.  R.
Oppenheimer fellowship  and by Department of Energy. Part of this
work was done at the Aspen Center for Physics, whose support is
acknowledged.

\eject

\figure{ Possible phase diagram for the  superconducting transition at $\alpha
> -1$. Reentrant evolution of the system from normal to superconducting to
normal state at $T_1$ and $T_2$ temperatures is shown. The minimal coupling
$g_{min}$  is required for a superconducting transition. The regime in which
Eq.(\ref{main}) is valid is violated  at $T \rightarrow 0$ and $T \sim W$  and
the dependence of $g$ on  $T$ is unknown. However we expect that as   $T
\rightarrow 0$ the critical coupling is  finite.}

\end{document}